\documentclass[prd,aps,showpacs,eqsecnum,floatfix,nofootinbib,preprint,tightenlines]{revtex4}
\usepackage{latexsym}
\usepackage{epsf}
\usepackage{graphicx}
\usepackage{slashed}
\usepackage{amsmath}
\usepackage{amssymb}

\newcommand{\pref}[1]{(\ref{#1})}

\newcommand{\mN}{M_N}

\newcommand{\ga}{g_A}
\newcommand{\Npp}{\mbox{$N\pi\pi$}\,\,}
\newcommand{\npp}{N\pi\pi}
\newcommand{\vecq}{\vec{q}}
\newcommand{\vecr}{\vec{r}}
\newcommand{\vecp}{\vec{p}}
\newcommand{\qr}{\vecq\cdot\vecr}

\newcommand{\Etot}{E_{\rm tot}}

\newcommand{\Epiq}{E_{\pi,{\vec{q}}}}
\newcommand{\Epir}{E_{\pi,{\vec{r}}}}

\newcommand{\nn}{\nonumber}
\newcommand{\cc}[1]{c_{#1}}
\newcommand{\qoe}{\frac{q^2}{\Epiq^2}}
\newcommand{\roe}{\frac{r^2}{\Epir^2}}
\newcommand{\qroe}{\frac{\qr}{\Epiq \Epir}}
\newcommand{\qqoe}{\frac{q^4}{\Epiq^4}}
\newcommand{\rroe}{\frac{r^4}{\Epiq^4}}
\newcommand{\qrqroe}{\frac{(\qr)^2}{\Epiq^2 \Epir^2}}
\newcommand{\qrqrqroe}{\frac{(\qr)^3}{\Epiq^3 \Epir^3}}

\begin{document}
\renewcommand{\thefootnote}{$*$}

\preprint{HU-EP-18/06}

\title{Three-particle $N\pi\pi$ state contribution to the 
nucleon two-point function in lattice QCD}

\author{Oliver B\"ar$^{a}$} 
\affiliation{$^a$Institut f\"ur Physik,
\\Humboldt Universit\"at zu Berlin,
\\12489 Berlin, Germany\\}

\begin{abstract}
The three-particle $\npp$ state contribution to the QCD two-point function of standard nucleon interpolating fields is computed to leading order in chiral perturbation theory. Using the experimental values for two low-energy coefficients the impact of this contribution on lattice QCD calculations of the nucleon mass is estimated. The impact is found to be at the per mille level at most and negligible in practice.
\end{abstract}

\maketitle

\renewcommand{\thefootnote}{\arabic{footnote}} \setcounter{footnote}{0}

\newpage
\section{Introduction}\label{Intro} 

In a recent paper \cite{Ce:2016ajy} (see also \cite{Ce:2017ndt,Giusti:2017ksp}) C\`{e}, Giusti and Schaefer proposed a factorization of the gauge field dependence of the fermion determinant in lattice QCD. Together with the factorization of the propagator \cite{Ce:2016idq} this allows the use of multi-level Monte-Carlo sampling methods in the calculation of correlation functions that suffer from large statistical uncertainties due to the signal-to-noise problem \cite{Parisi:1983ae,Lepage:1989hd}.
First tests of this proposal are encouraging \cite{Ce:2016ajy}, and one can expect significantly reduced statistical errors in lattice computations of many phenomenologically interesting observables.     

Smaller statistical errors imply that more sources of systematic uncertainty need to be considered that used to be negligible before.
An example is the excited state contamination in correlation functions that are measured to calculate physical observables. With the up and down quark masses at their physical value multi-particle states with additional pions are a non-negligible source of systematic uncertainty. Recent calculations \cite{Bar:2016uoj,Bar:2016jof} in chiral perturbation theory (ChPT) \cite{Weinberg:1978kz,Gasser:1983yg,Gasser:1984gg} suggest a 5-10\% overestimation of various nucleon observables (nucleon charges, moments of structure functions) by lattice calculations, caused by the two-particle nucleon-pion ($N\pi$) state contribution to 3-point (pt) correlation functions. 

The $N\pi$ contribution can be expected to be the dominant multi-hadron state contribution at large time separations, but there are other contributions as well. The $\Delta\pi$ contribution to the effective nucleon mass and the axial charge was calculated  within heavy baryon ChPT in Refs.\ \cite{Tiburzi:2009zp,Tiburzi:2015tta} and found to be significantly smaller then the $N\pi$ contribution. Also the  $\npp$ contribution is expected to be much smaller since it is a three-particle-state contribution. However, it is unknown how large it actually is, and simple estimates based on the expected O($1/L^6)$ suppression of the finite volume matrix elements can be quite misleading. The multi-particle-state contribution is not a finite volume effect that vanishes in the infinite volume limit; it is a non-vanishing cumulative contribution caused by a large number of states even in volumes of moderate size  with $M_{\pi}L=4$, for instance. 

Here we report the results of a ChPT calculation of the $\npp$ contribution to the nucleon 2-pt function and the nucleon effective mass. The computation is analogous to the calculation of the $N\pi$ contribution in Ref.\ \cite{Bar:2015zwa}. To LO in the chiral expansion the $\npp$ contribution depends on two LO low-energy coefficients (LECs) only, the nucleon axial charge and the pion decay constant. Taking the known experimental values as input the impact of the $\npp$ contribution on lattice calculations of the nucleon mass can be estimated. We find it to be at the per mille level for source-sink separations of 1.2 fm and larger. This is indeed very small and lattice data with statistical error at sub per mille level are needed to be sensitive to the $\npp$ contribution. Since statistical errors at present are much larger the $\npp$ contribution can be safely ignored. Although here we consider the 2-pt function only it seems safe to assume that the $\npp$ contribution to nucleon 3-pt functions is similarly suppressed and negligible too. 

The calculation of the $\npp$ contribution is very similar to the analogous one in Ref.\ \cite{Bar:2015zwa} of the $N\pi$ contribution. We also refer to recent reviews \cite{Bar:2017kxh,Bar:2017gqh} for the general strategy and setup of this kind of calculations. Here we will be brief and focus on the particular aspects of the three-particle $\npp$ contribution. 

\section{$\mathit{N}\mathbf{\pi\pi}$ contribution to the nucleon 2-pt function}\label{secQCD} 

\subsection{Setup}\label{ssect:setup}
We consider QCD with equal up and down quark masses. The spatial volume is assumed to be finite with spatial extent $L$, and periodic boundary conditions are imposed in each direction. The euclidean time extent is taken infinite, for simplicity. 

We are interested in the 2-pt function of nucleon interpolating fields $N,\overline{N}$ with positive parity,
\begin{equation}\label{Def2pt}
G_{\rm 2pt}(t) = \int d^3x \,\Gamma_{\alpha\beta} \langle N_{\beta}(\vec{x},t) \overline{N}_{\alpha}(\vec{0},0)\rangle\,,
\end{equation}
with $\Gamma=(1+\gamma_0)/4$. We assume  $N, \overline{N}$ to be given by the standard local 3-quark operators without derivatives \cite{Ioffe:1981kw}. We can also allow for smeared quark fields provided a) the smearing method is compatible with chiral symmetry\footnote{Familiar examples are Gaussian or exponential smearing \cite{Gusken:1989ad,Gusken:1989qx,Alexandrou:1990dq} and the gradient flow \cite{Luscher:2013cpa}, for instance.} and  b) the ``size'' of the smeared interpolating fields is much smaller than the Compton wavelength of the pion. If that is the case smeared interpolating fields are mapped to the same point-like expressions in ChPT as their local counterparts \cite{Bar:2013ora,Bar:2015zwa}. 

Performing the usual spectral decomposition in \pref{Def2pt} for large euclidean times $t\gg0$ the dominant contribution stems from the single-nucleon state $|N(\vec{p}=0)\rangle$ describing the nucleon at rest,
\begin{equation}\label{spcontr}
G_{\rm 2pt}^N(t)=
\frac{1}{2M_{N}}\;|\langle 0|N(0)|N(\vec{p}=0)\rangle|^2 e^{-M_N t } \,,
\end{equation}
dropping off with the nucleon mass $M_N$. The interpolating field excites other states with the quantum numbers of the nucleon as well. For small physical pion masses the dominant multi-hadron states are those containing the nucleon and additional light pions. For the three-particle $\npp$ state contribution we find \begin{eqnarray}\label{nppcontr}
G_{\rm 2pt}^{\npp}(t)&=&\frac{1}{L^6}\;\sum_{\vec{q},\vec{r}}\frac{1}{8E_{N,\vec{p}} E_{\pi,\vec{q}} E_{\pi,\vec{r}}}\,
|\langle 0|N(0)|N(\vec{p}) \pi(\vec{q})\pi(\vec{r})\rangle|^2 e^{-E_{\rm tot}t}\,.
\end{eqnarray}
The spatial momenta $\vecq,\vecr$ refer to the momenta of the two pions, the nucleon momentum $\vecp$ is fixed by momentum conservation, $\vecp = -\vecq-\vecr$. The energy $\Etot$ denotes the total energy of the 3-particle state. For weakly interacting pions it approximately equals the sum of the individual hadron energies. The sum in \pref{nppcontr} runs over all momenta compatible with the periodic boundary conditions that we have assumed to be imposed, i.e.\ 
\begin{equation}\label{pimomenta}
\vec{q}=\frac{2\pi}{L}\vec{n}_q\,,\qquad \vec{r}=\frac{2\pi}{L}\vec{n}_r\,,
\end{equation}
with the two vectors $\vec{n}_q,\vec{n}_r$ having integer valued components. The absolute values of the pion momenta can be labelled by integers $n_q,n_r$, defined according to 
\begin{equation}\label{Defnq}
|\vec{q}|=\frac{2\pi}{L}\sqrt{n_q}\,, \quad n_q={n}_{q,x}^2+{n}_{q,y}^2+{n}_{q,z}^2\,,
\end{equation}
and analogously for $\vec{r}$.

\subsection{Chiral Perturbation Theory}\label{ssect:general}
Correlation functions like the 2-pt function of the previous section can be computed perturbatively in ChPT, provided the time separation $t$ is sufficiently large such that the correlation function is dominated by the light pions. Here we employ the covariant formulation of SU(2) Baryon ChPT (BChPT) to leading order in the chiral expansion \cite{Gasser:1987rb,Becher:1999he}. To this order the effective  Lagrangian is the sum of two parts, ${\cal L}_{\rm eff}={\cal L}_{N\pi}^{(1)} + {\cal L}_{\pi\pi}^{(2)}$. The latter one, ${\cal L}_{\pi\pi}^{(2)}$, denotes the two-flavor mesonic chiral Lagrangian to LO \cite{Gasser:1983ky}. The first part, ${\cal L}_{N\pi}^{(1)}$, contains the nucleon fields and their coupling to the pions. We assume isospin symmetry, thus the effective theory contains three mass degenerate pions $\pi^a, a=1,2,3$, and the nucleon doublet $\Psi=(p,n)^T$ with the fields for the mass degenerate proton and neutron. Expanding the chiral effective Lagrangian and keeping interaction terms with up to two pion fields we find the interaction Lagrangian to be given by\footnote{We work in Euclidean space time.}
\begin{equation}\label{intlagrangian}
{\cal L}_{{\rm int}} =\frac{i\ga}{2f}\overline{\Psi}\gamma_{\mu}\gamma_5\sigma^a\Psi \partial_{\mu}\pi^a 
- \frac{i}{4f^2}\epsilon^{abc}\pi^a\partial_{\mu}\pi^b \overline{\Psi}\gamma_{\mu}\sigma^c\Psi 
\,.
\end{equation}
It involves the LO LECs $\ga$ and $f$, the chiral limit values of the axial charge and the pion decay constant. Since we work to LO it is consistent to replace these by their experimental values. With our conventions these are $\ga =1.2727$ and $f_{\pi}=92.4$MeV.

In a similar fashion we expand the ChPT expressions for the nucleon interpolating fields derived in Ref.\ \cite{Wein:2011ix}. To LO and up to two pion fields we obtain
\begin{eqnarray}
N& = & \tilde{\alpha} \left(\Psi + \frac{i}{2f} \pi^a\sigma^a \gamma_5\Psi-\frac{1}{8f^2}\pi^a\pi^a \Psi    \right)\,,\label{Neffexp}\\
\overline{N} & = & \tilde{\beta}^*\left(\overline{\Psi} + \frac{i}{2f}\overline{\Psi}\gamma_5\sigma^a\pi^a -\frac{1}{8f^2}\overline{\Psi}\pi^a\pi^a\right)\,.\label{Nbareffexp}
\end{eqnarray}
Here $ \tilde{\alpha}, \tilde{\beta}$ are the LECs associated with the interpolating fields. If the same interpolating fields are used at source and sink the LECs are the same, $\tilde{\alpha} = \tilde{\beta}$.

The expressions in \pref{Neffexp}, \pref{Nbareffexp} are the effective fields for both point-like and smeared fields provided the smearing procedure is compatible with chiral symmetry and ``smearing radii'' small compared to the pion Compton wavelength. Any differences between point like and smeared interpolating fields are encoded in different values for the LECs $\tilde{\alpha},\tilde{\beta}$ only. For physical pion masses the pion Compton wavelength is about 1.4 fm, thus one expects the expressions in \pref{Neffexp} to be valid for smearing radii of a few tenths of a fermi.
 
 \subsection{\label{sseceffFields} The $N\pi\pi$ contribution in the 2-pt function}
 
\begin{figure}[tbp]
\begin{center}
\includegraphics[scale=0.5]{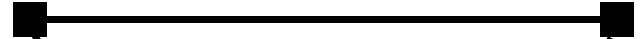}
\caption{Leading Feynman diagram with a single-nucleon-state contribution to the  2-pt function. Squares represent the nucleon interpolating fields at times $t$ and $0$ and the solid line depicts the nucleon propagator.}
\label{fig:Ndiag}
\end{center}
\end{figure}

With the expressions \pref{intlagrangian} - \pref{Nbareffexp} it is straightforward to compute the 2-pt function perturbatively. The first contribution stems from the Feynman diagram in fig.\ \ref{fig:Ndiag}. It is essentially the nucleon propagator and leads to the leading single-nucleon-state contribution in the 2-pt function,
\begin{equation}\label{SingleNucl}
G^{N}_{{\rm 2pt}} = \tilde{\alpha}\tilde{\beta}^* e^{-M_N t}\,. 
\end{equation} 
In case of the same interpolating fields at source and sink we can compare this expression with eq.\ \pref{spcontr} and find the relation between the LEC $|\tilde{\alpha}|$ and the matrix element in  \pref{spcontr}.

\begin{figure}[tbp]
\begin{center}
\includegraphics[scale=0.5]{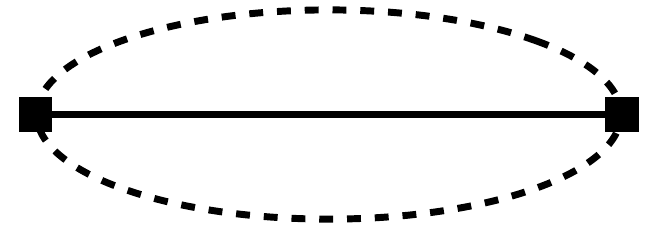}\hspace{0.3cm}\includegraphics[scale=0.5]{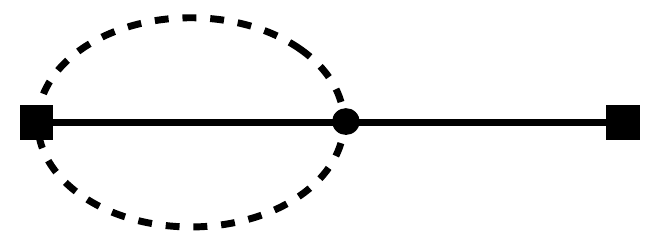}\hspace{0.3cm}\includegraphics[scale=0.5]{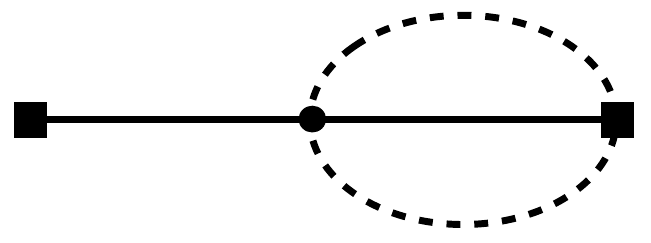}\hspace{0.3cm}\includegraphics[scale=0.5]{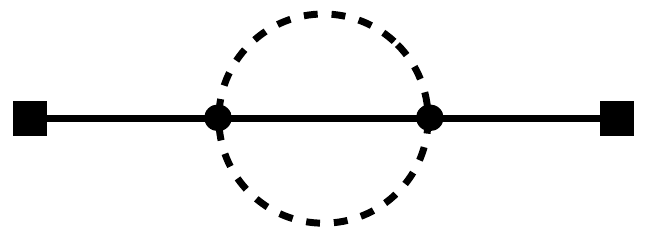}\\
a)\hspace{3cm} b)\hspace{3cm} c)\hspace{3cm} d)\\[3ex]
\includegraphics[scale=0.5]{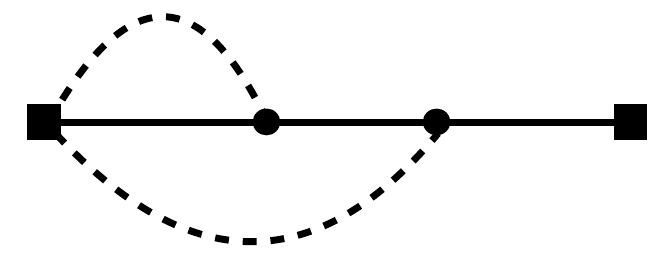}\hspace{0.3cm}\includegraphics[scale=0.5]{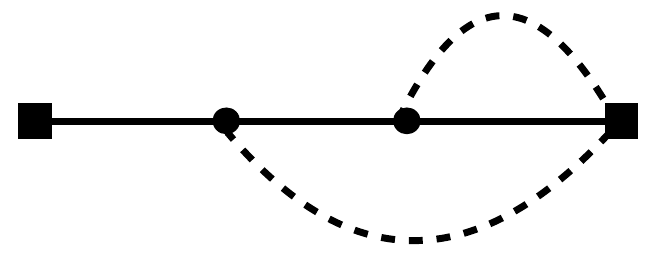}\hspace{0.3cm}\includegraphics[scale=0.5]{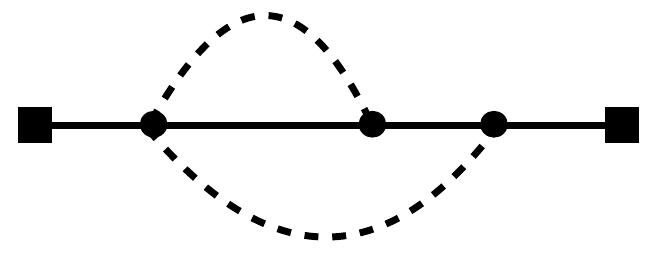}\hspace{0.3cm}\includegraphics[scale=0.5]{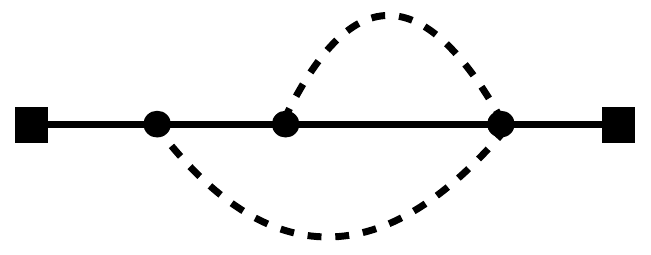}\\
e)\hspace{3cm} f)\hspace{3cm} g)\hspace{3cm} h) \\[3ex]
\includegraphics[scale=0.5]{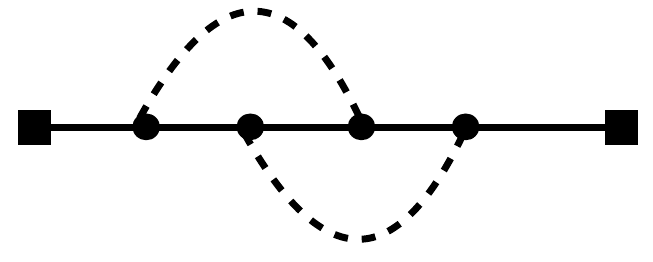}\hspace{0.3cm}\includegraphics[scale=0.5]{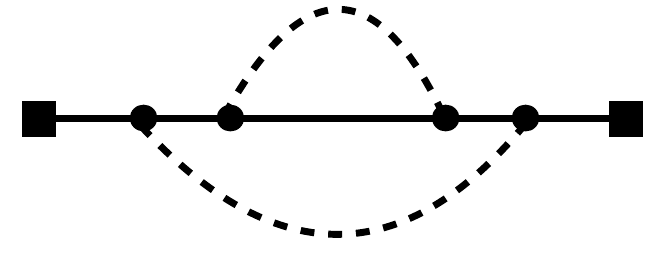}\hspace{0.3cm}\includegraphics[scale=0.5]{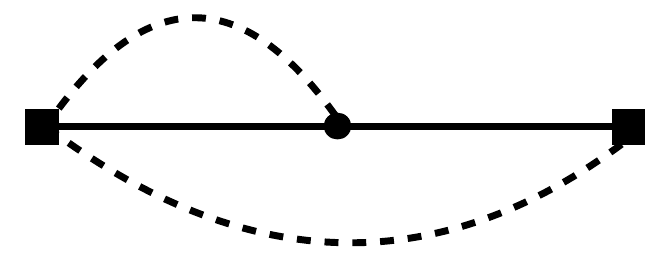}\hspace{0.3cm}\includegraphics[scale=0.5]{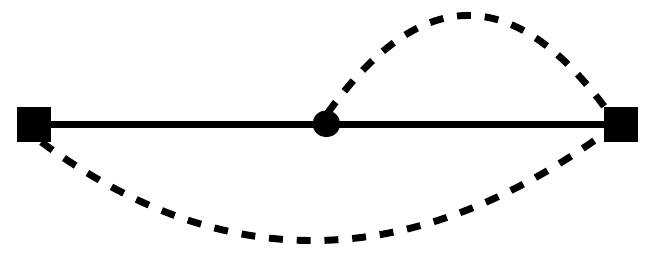}\\
i)\hspace{3cm} j)\hspace{3cm} k)\hspace{3cm} l)\\[3ex]
\includegraphics[scale=0.5]{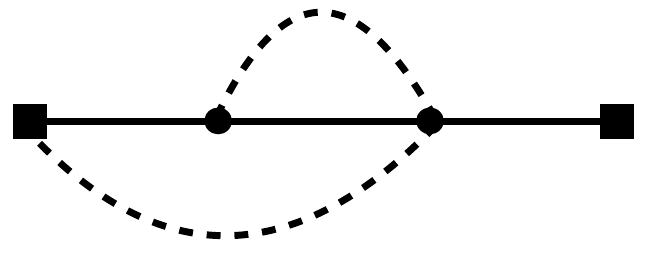}\hspace{0.3cm}\includegraphics[scale=0.5]{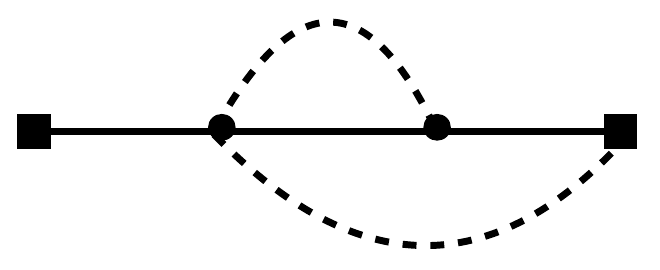}\hspace{0.3cm}\includegraphics[scale=0.5]{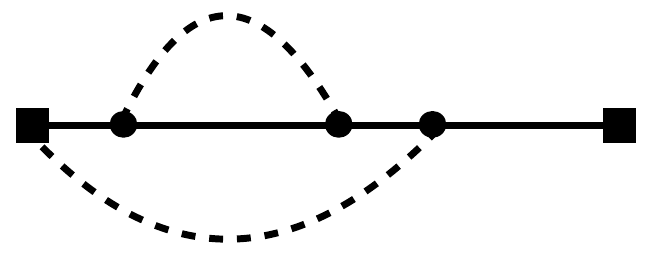}\hspace{0.3cm}\includegraphics[scale=0.5]{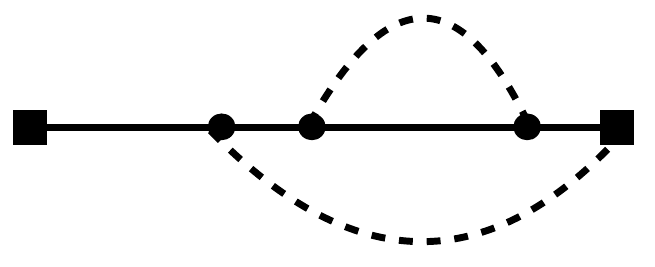}\\
m)\hspace{3cm} n)\hspace{3cm} o)\hspace{3cm} p)\\[3ex]
\includegraphics[scale=0.5]{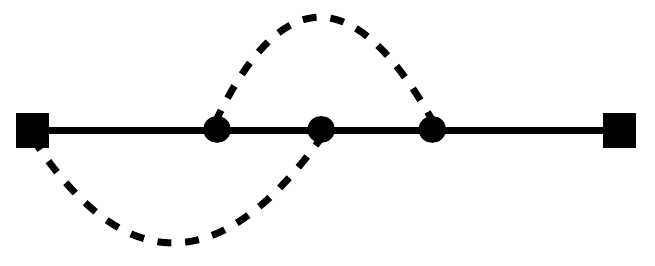}\hspace{0.3cm}\includegraphics[scale=0.5]{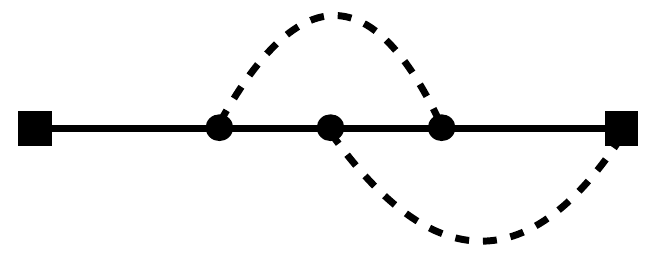}\hspace{0.3cm}\includegraphics[scale=0.5]{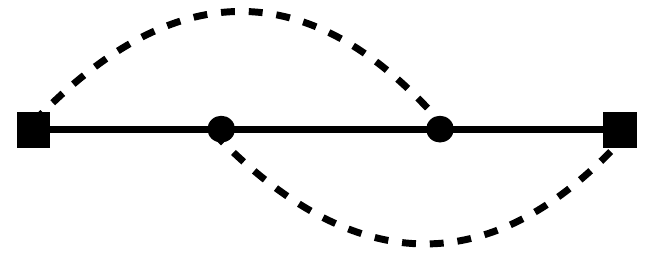} \\
q)\hspace{3.3cm} r)\hspace{3.3cm} s)
\caption{Feynman diagrams for the  2-pt function with an $N\pi\pi$ contribution. Circles represent a vertex insertion at an intermediate space time point, and an integration over this point is implicitly assumed. The dashed lines represent a pion propagator. }
\label{fig:Npipidiags}
\end{center}
\end{figure}

The leading contribution to the three-particle $N\pi\pi$ contribution $G^{N\pi\pi}_{{\rm 2pt}}$ stems from the  diagrams in fig.\ \ref{fig:Npipidiags}. 
Although 2-loop diagrams their $N\pi\pi$ contribution does not involve any summation over some undetermined loop momentum, so we essentially perform a tree-level calculation. 


In order to present the results we introduce some useful notation.  
The \Npp contribution to the 2-pt function is of the following general form:
\begin{equation}\label{nppcontrgeneral}
C_{\rm 2pt}^{N\pi\pi}(t) = \tilde{\alpha}\tilde{\beta}^*\sum_{\vec{q},\vec{r}} c_{\vecq,\vecr}\, e^{-\Etot t}\,.
\end{equation}
To the order we are working the energy $\Etot$ is just the sum of the individual hadron energies,
\begin{eqnarray}
\Etot &=& E_{{N},\vecp} +E_{\pi,\vecq}+E_{\pi,\vecr}\,,\nn\\
E_{{N},\vecp}& = &\sqrt{\vecp^{\,2} +M_N^2}\,,\nn\\
E_{\pi,\vecq} & =&  \sqrt{\vecq^{\,2}+M_{\pi}^{2}}\,.\label{energies}
\end{eqnarray}
The coefficients $c_{\vecq,\vecr}$ in \pref{nppcontrgeneral} are the non-trivial result of the calculation presented here. 
It is convenient to write them as a product of some universal factor and some ``reduced'' coefficients $C_{\vecq,\vecr}$,
\begin{equation}\label{defredCoeff}
c_{\vecq,\vecr} = \frac{3}{128 (fL)^4 (\Epiq L )(\Epir L)} C_{\vecq,\vecr}\,.
\end{equation}
The fraction on the right hand side contains the expected $1/L^6$ dependence of a 3-particle state in a finite spatial volume, cf.\ eq.\ \pref{nppcontr}. The inverse mass dimensions combine with the pion decay constant $f^4$ and the pion energies to the dimensionless combination in the denominator. The factor 3 in the numerator simply counts the number of pions in the 2-flavor theory. The numerical factor 128 in the denominator is chosen such that the reduced coefficient $C_{0,0}$ of the state with all three particles at rest is equal to one in the infinite nucleon mass limit, see below.

The reduced coefficients are dimensionless and depend on (ratios of) the momenta, energies and masses of the pions and the nucleon. In addition they depend on the dimensionless LEC $g_A$. The results for the coefficients $C_{\vecq,\vecr}$ are quite cumbersome except for some special cases where one or all three particles are at rest. For this reason we perform the non-relativistic expansion of the nucleon energy and keep only the first two terms in this expansion. The truncation error caused by this expansion is expected to be much smaller than the higher order corrections to our LO results, so for our purposes this non-relativistic approximation should be more than sufficient. Explicitly, we expand
\begin{equation}
E_{N,\vec{p}}= M_N +\frac{\vecp^{\,2}}{2M_N} +{\rm O}\left(\frac{1}{M_N^2}\right)
\end{equation}
in the coefficients $C_{\vecq,\vecr}$ and drop all contributions of O($1/M_N^2)$ and higher. The results we obtain this way are given according to 
\begin{equation}\label{DefNRCoeff}
C_{\vecq,\vecr} = C^{\infty}_{\vecq,\vecr} + \frac{\Epiq+\Epir}{M_N} C^{\rm corr}_{\vecq,\vecr}\,.
\end{equation}
The coefficients  $C^{\infty}_{\vecq,\vecr}, C^{\rm corr}_{\vecq,\vecr}$ depend only on the momenta, energies and mass of the pion,  and on the LEC $g_A$. Thus they are finite in the limit $M_N\rightarrow \infty$, and $C^{\infty}_{\vecq,\vecr}$ is the infinite-nucleon-mass limit of the coefficients. 

For later reference we quote the result for the simplest case separately, namely the one with all three particles at rest. This state has the lowest energy of all $\npp$ states, and for this contribution we find
\begin{equation}\label{resC00}
C_{0,0}^{\infty} = 1\,,\quad C_{0,0}^{\rm corr} = \frac{g_A^2}{2}-g_A\,.
\end{equation}
As mentioned earlier, the result $C_{0,0}^{\infty} = 1$ is a consequence of the particular normalization in \pref{defredCoeff}, i.e.\ the presence of the numerical prefactor 3/128 in the universal factor. This definition ensures that the reduced coefficients are of O(1).\footnote{In Ref.\ \cite{Bar:2012ce} the three-particle $\pi\pi\pi$ contribution in the 2-pt function of the pseudo-scalar density was computed. In that case the prefactor equals 45/512. This is about 4 times larger than 3/128, but the overall size of the three-particle-state contributions is comparable.}

Rotation invariance implies that the result for the general case with non-vanishing pion momenta $\vecq$ and $\vecr$ can depend only on $q^2=\vecq\cdot\vecq$, $r^2=\vecr\cdot\vecr$ and $\vecq\cdot \vecr$ (the pion energies are determined by eq.\ \pref{energies}).
In terms of these variables we find the result 
\begin{eqnarray}\
C_{\vecq,\vecr}^{\infty} & = & 1 + 2 \left(\frac{\Epiq-\Epir}{\Epiq+\Epir} \right)^2\nn \\ 
& & \phantom{1} - 2g_A^2\frac{\Epiq^2-6\Epiq\Epir+\Epir^2}{(\Epiq+\Epir)^2}\,\frac{\qr}{\Epiq\Epir}\nn\\
& & \phantom{1} - 4g_A^4\frac{\Epiq\Epir}{(\Epiq+\Epir)^2}\,\left(\frac{\qr}{\Epiq\Epir}\right)^2\nn\\
& & \phantom{1} +g_A^4  \,\frac{3\Epiq^2+2\Epiq\Epir+3\Epir^2}{(\Epiq+\Epir)^2}\,\frac{q^2 r^2}{\Epiq^2\Epir^2}\,.\label{resNRLimCqr}
\end{eqnarray}
Setting both pion momenta equal to zero eq.\ \pref{resNRLimCqr} reproduces $C_{0,0}^{\infty}$ in \pref{resC00}.

The result for the O($1/M_N$) correction $C_{\vecq,\vecr}^{\rm corr}$ is somewhat lengthier and given in appendix \ref{appA}.
We will see in the next section that it amounts to a thirty percent correction to the leading result based on the coefficients \pref{resNRLimCqr}.

The non-relativistic expansion in \pref{DefNRCoeff} simplifies significantly the calculation and the results for the coefficients $C_{\vecq,\vecr}$. In particular, diagrams k) - r) in fig.\ \ref{fig:Npipidiags} start contributing at O($1/M_N$) only. These diagrams have a single $N\pi$-vertex from the interpolating fields in common. This vertex involves a single $\gamma_5$ matrix, thus it is O($1/M_N$) suppressed \cite{Krause:1990xc}. Diagram s) involves two of these vertices. It is found to be of O($1/M_N^2$) and does not contribute to the order we are working here.
 
\subsection{Impact on lattice calculations}\label{secNumEst}

Let us estimate the impact of the $\npp$ state contribution on lattice calculations of the nucleon mass. The nucleon mass is usually obtained from the effective mass, the negative time derivative of $\ln G_{\rm 2 pt}(t)$. With \pref{SingleNucl} and \pref{nppcontrgeneral} we obtain
\begin{equation}\label{DefMeff}
M_{{\rm eff}}(t)  =  M_N \left[ 1+ \sum_{\vec{q},\vec{r}} d_{\vec{q},\vec{r}} \,e^{-(E_{{\rm tot}} -\mN) t}\right],
\end{equation}
 with the coefficients $d_{\vec{q},\vec{r}} $ being related to the previously defined ones according to 
\begin{equation}
d_{\vec{q},\vec{r}}  =  c_{\vec{q},\vec{r}}\left[\frac{E_{{\rm tot}}}{M_N}-1\right].
\end{equation}
Note that the effective mass does not depend on the LO LECs $\tilde{\alpha},\tilde{\beta}$ associated with the nucleon interpolating fields. These are overall factors in the 2-pt function and drop out in the effective mass. Thus, the $\npp$ contribution with coefficients $d_{\vec{q},\vec{r}}$ to LO is the universal contribution valid for both local and smeared interpolating fields. This universality property will be lost only at higher order in the chiral expansion where additional LECs will enter the chiral expressions for the nucleon interpolating fields and the final result for the effective mass.

To LO the $\npp$ contribution depends on the pion and nucleon masses, the spatial extent $L$, and on the LECs $g_A$ and $f_{\pi}$. These LECs are experimentally very well known \cite{Patrignani:2016xqp}. Using the experimental values 1.27 and 93 MeV as input in the ChPT result we can estimate the impact of the three-particle $\npp$ states in lattice calculations of the nucleon mass. The only free parameter we need to fix is $L$. If not specified otherwise we do this by choosing the common value $M_{\pi}L=4$ for physical pion mass $M_{\pi}=140$ MeV. The nucleon mass is given by $M_N=940$ MeV. 

\begin{figure}[tbp]
\begin{center}
\includegraphics[scale=0.9]{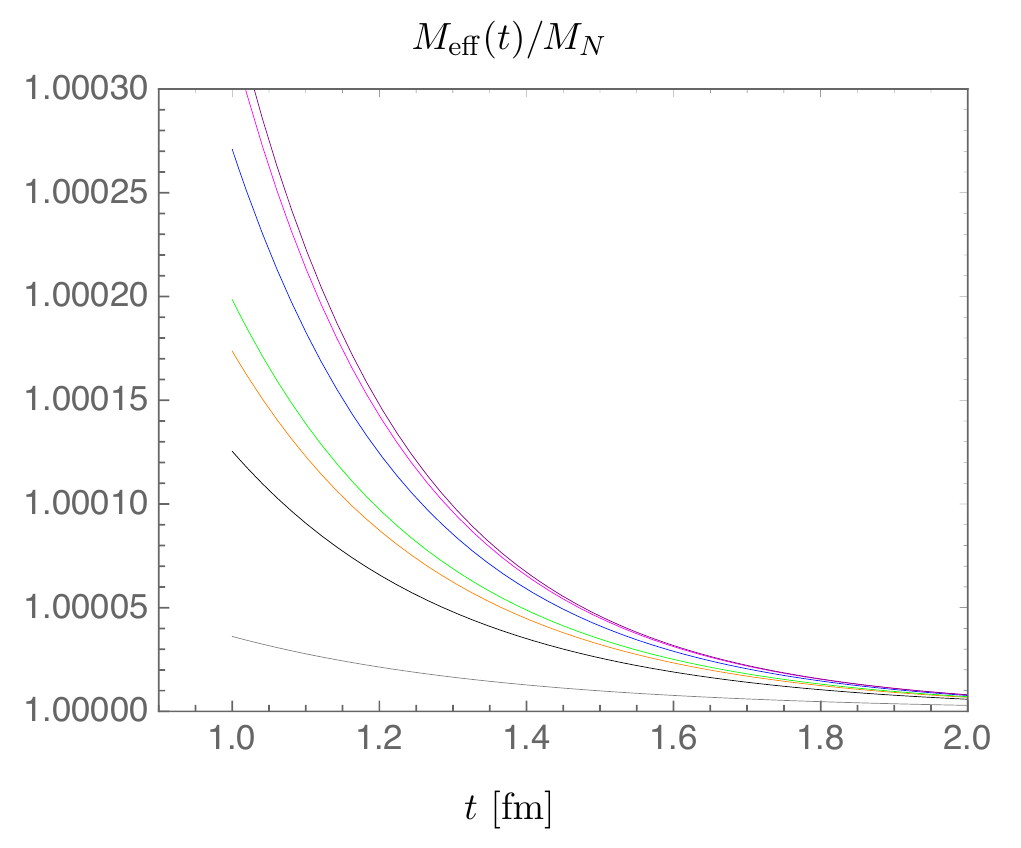}\\[-1ex]
\caption{The ratio $M_{\rm eff}(t)/M_N$ for seven lower bounds, starting from  $n_{\rm max}=1$ (gray line) to $n_{\rm max}=8$ (purple line). The spatial extent satisfies $M_{\pi}L=4$.}
\label{fig:res1}
\end{center}
\end{figure}

Fig.\ \ref{fig:res1} shows the ratio $M_{{\rm eff}}(t) / M_N$ as a function of euclidean time $t$. Without any excited state contribution this ratio would be constant and equal to 1. The figure shows the deviations from 1 due to the $\npp$ contribution with various upper bounds for the two pion momenta.
Displayed are the results for $n_q,n_r \le n_{\rm max}$ for $n_{\rm max}=1$ (gray curve) up to $n_{\rm max}= 8$ (purple curve).\footnote{See appendix \ref{appB} for some details.} These numbers can be translated into values for the expansion parameter $p_n/\Lambda_{\chi}$ of finite volume ChPT, with $\Lambda_{\chi}$ being the chiral scale usually identified with $4\pi f_{\pi}\approx 1.2$ GeV \cite{Colangelo:2003hf}. 
For $M_{\pi}L=4$ the values $n_{\rm max} = 2$ and 5 correspond to $p_{n_{\rm max}}/\Lambda_{\chi}\approx 0.3$ and 0.45, respectively \cite{Bar:2015zwa}. 

For ChPT to give trustworthy results the time separation $t$ needs to be sufficiently large such that the low-momentum $\npp$ contribution dominates the high-momentum contribution which is not well captured by ChPT. In that case the latter can be ignored with a small truncation error. Looking at fig.\ \ref{fig:res1} one can expect this to be the case for $t$ about 1.2 fm and larger. There is no need to be more precise here, since the $\npp$ contribution is of order $10^{-4}$ for these times. Even if we allow for a generous factor 2 due to the higher momentum states and an additional factor 2 for the higher order chiral corrections the $\npp$ contribution to the effective mass is $10^{-3}$ at most. Consequently, the $\npp$ contribution can be safely neglected unless  lattice data with statistical errors at the sub per mille level are available.        

\begin{figure}[tbp]
\begin{center}
\includegraphics[scale=0.9]{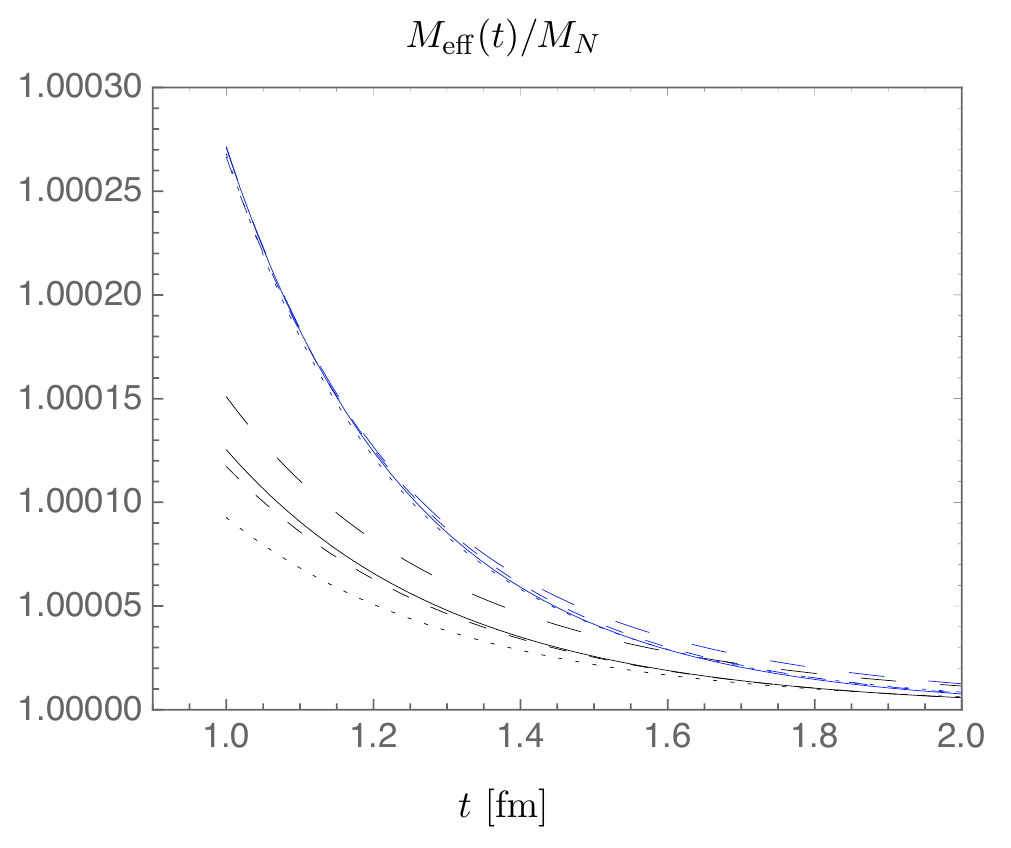}\\[-1ex]
\caption{The ratio $M_{\rm eff}(t)/M_N$ for four different spatial volumes: $M_{\pi}L=2.5$ (long-dashed), $M_{\pi}L=3$ (dotted), $M_{\pi}L=4$ (solid) and $M_{\pi}L=5$ (short-dashed). The upper bounds on the pion momenta are $n_{\rm max}=1,1,2$ and 4 (black) and $n_{\rm max}=2,3,5$ and 8 (blue), respectively. These values correspond to approximately the same bounds for the momenta in physical units. For the larger of the two bounds the $\npp$ contribution is essentially volume independent.}
\label{fig:res2}
\end{center}
\end{figure}

The two-particle $N\pi$ contribution was observed to be essentially independent of the spatial volume provided the number of states taken into account is appropriately adjusted as $L$ is changed. The same holds for the three-particle $\npp$ contribution, as 
fig.\ \ref{fig:res2} shows. The $\npp$ contribution is displayed for various spatial volumes: $M_{\pi}L=2.5$ (long-dashed lines), $M_{\pi}L=3$ (dotted lines), $M_{\pi}L=4$ (solid lines) and $M_{\pi}L=5$ (short-dashed lines). The upper bounds for the pion momenta correspond to $n_{\rm max}=1$ and 2, $n_{\rm max}=1$ and 3, $n_{\rm max}=2$ and 5  and $n_{\rm max}=4$ and 8, respectively. These different values correspond to approximately the same bounds for the momenta in physical units, namely the values $p_{n_{\rm max}}/\Lambda_{\chi}=0.3$ and 0.45.\footnote{See also table 1 in Ref.\ \cite{Bar:2017kxh}.} 
While there is still some FV dependence visible for the smaller momentum bound (black lines), the differences between curves for the larger momentum bound (blue lines) are very small. Comparing the results for the largest two volumes the differences are barely visible and we may conclude that the results for volumes with $M_{\pi}L=4$ capture the infinite-volume result already very well.
Note, however, that the weak volume dependence will be lost if $n_{\rm max}$ is kept fixed as the volume is changed. In that case a strong volume dependence, stemming essentially from the $1/L^6$ dependence in the coefficients $c_{\vecq,\vecr}$, is clearly visible in the $\npp$ contribution when $L$ is changed.

Although small the $\npp$ contribution is much larger than naive estimates may suggest. For example, the contribution \pref{resC00} of the lowest $\npp$ state with all three particles at rest results in a mere $1.37\times 10^{-6}$ to the ratio $M_{\rm eff}(t)/M_N$ at $t=1.2$ fm for $M_{\pi}L=4$. This is a factor 90 times smaller than the total contribution of all $\npp$ states with $n_{\rm max} = 5$ (cf.\ solid blue line in fig.\ \ref{fig:res2}). The contribution of the single $\npp$ state drops even further to $3.59 \times 10^{-7}$ for $M_{\pi}L=5$, and the factor 90 increases to about 350. These substantial factors stem from the large number of \Npp states that contribute significantly to the sum in \pref{nppcontrgeneral} already at moderatly large spatial volumes.

\begin{figure}[tbp]
\begin{center}
\includegraphics[scale=0.9]{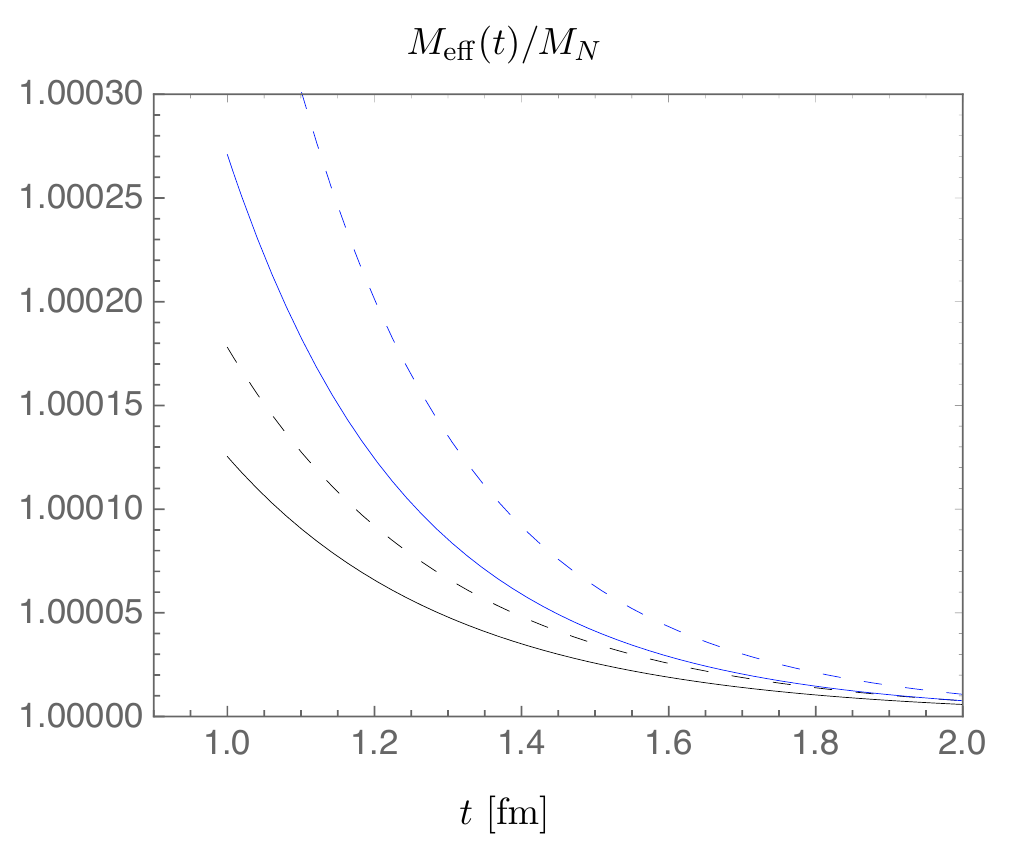}\\[-1ex]
\caption{The ratio $M_{\rm eff}(t)/M_N$ for $M_{\pi}L=4$ and upper bounds $n_{\rm max}=2$ (black) and $n_{\rm max}=5$ (blue). Shown are the results for the \Npp correction with (solid lines) and without (dashed lines) the O($1/M_N)$ correction $C^{\rm corr}_{\vecq,\vecr}$.}
\label{fig:res3}
\end{center}
\end{figure}

Figure \ref{fig:res3} illustrates the size of the O($1/M_N$) correction in our results. Plotted are, for $M_{\pi}L=4$,  the $\npp$ contribution with (solid lines) and without (dashed lines) the correction term $C^{\rm corr}_{\vecq,\vecr}$ in the coefficients, cf.\ \pref{DefNRCoeff}. The results are shown for two different upper momentum bounds, $n_{\rm max}=2$ (black lines) and $n_{\rm max}=5$ (blue lines). The O($1/M_N$) correction amounts in an approximately 30\% decrease of the infinite-nucleon-mass result. The size of this correction agrees with naive expectations, but the sign of the correction is a priori not known.
 
In this paper we are mainly interested in the total \Npp state contamination in the nucleon 2-pt fuction, which is the cumulative contribution of a substantial number of low-momentum \Npp states. However, it is worth mentioning that the generalized eigenvalue method \cite{Luscher:1990ck} allows to determine ratios of matrix elements with individual \Npp states and the single nucleon state, and these ratios can be directly compared with the ChPT results for the coefficients $c_{\vecq,\vecr}$ in \pref{nppcontrgeneral}. 

In a recent paper \cite{Lang:2016hnn}  a correlation matrix involving five interpolating fields was computed and analyzed. The calculation is based on the PACS-CS ensemble of gauge configurations with $M_{\pi}\approx 156$ MeV and $L\approx 2.9$ fm \cite{Aoki:2008sm}, and the numerical result
\begin{equation}\label{c00Lat}
\sqrt{c_{0,0}^{\rm Lat}} = 0.07\pm0.04
\end{equation}
is found \cite{SPprivcomm}.\footnote{With the notation of Ref.\ \cite{Lang:2016hnn} we find $|Z^{n=2}_{i=6}/Z^{n=1}_{i=6}|^2 = c^{\rm Lat}_{0,0}$.} The ChPT result \pref{defredCoeff}, \pref{resC00} gives
\begin{equation}
\sqrt{c_{0,0}^{\rm ChPT}} = 0.036
\end{equation}
and agrees with \pref{c00Lat} within the large numerical error. Although encouraging the comparison is not unproblematic in this particular case: The smearing radii of the smeared interpolating fields used in \cite{Lang:2016hnn} exceed 0.5 fm and are uncomfortably large, while the value $M_{\pi}L\approx 2.2$ is rather small for ChPT to be in the $p$-regime. Thus, one would not be too surprised to see a sizable discrepancy between the numerical and the ChPT result if the statistical error were smaller. 
 
\section{\label{secConcl} Conclusions}
The LO ChPT results for the \Npp contribution to the nucleon 2-pt function are very small. Unless lattice data with sub per mille precision are available the \Npp contribution to the effective nucleon mass can be safely ignored. The only contribution relevant in practice will be the two-particle $N\pi$ contribution \cite{Tiburzi:2015tta,Bar:2015zwa}. Although small as well it is about a hundred times larger than the \Npp  contribution and affects the effective mass at the percent level. 

Based on the results shown here we can expect the $\npp$ contribution to nucleon 3-pt functions to be small and negligible as well. The two-particle $N\pi$ contribution in the determination of various nucleon charges and moments of structure functions was shown to be at the 5-10\% level for source-sink separations of $2$ fm \cite{Bar:2016uoj,Bar:2016jof}. If we assume the same suppression factor that we found in the 2-pt function we can estimate the $\npp$ contribution in calculations of the nucleon charges and moments to be at the per mille level. Once again this is completely negligible in practice at the moment. 

The excited-state contribution in the plateau estimate of the axial charge is rather peculiar.  
Existing lattice data underestimate the experimental value for source-sink separations less than about 1.5 fm, while for 2 fm and larger 
the LO ChPT result for the $N\pi$ contribution predicts an overestimation by lattice calculations. This behavior is not seen for other observables like the average quark momentum fraction or the helicity moment \cite{Bar:2017kxh,Bar:2017gqh}. The most likely explanation are excited states other than the two-particle $N\pi$ states that provide an additional negative excited-state contribution to the plateau estimate. Whether this is correct and, if correct, which states these are is unknown to date, but based on the results presented here the three-particle \Npp states can essentially be ruled out as potential candidates. 

\vspace{2ex}
\noindent {\bf Acknowledgments}
\vspace{2ex}

I thank S.\ Prelovsek for discussion and correspondence.
This work was supported by the German Research Foundation (DFG), Grant ID BA 3494/2-1.
\vspace{3ex}

\begin{appendix}

\section{The $\mathbf{O( 1/{M_N})}$ term in the reduced coefficient}
\label{appA}
The O$(1/M_N)$ correction $C^{\rm corr}_{\vecq,\vecr}$ is defined in \pref{DefNRCoeff}. 
It depends on the absolute values of the pion momenta, $q^2,r^2$ and the scalar product $\qr$. The pion energies $\Epiq,\Epir$ are given in terms of these according to \pref{energies}. The calculation of the diagrams in fig.\ \ref{fig:Npipidiags} yields the following result:\footnote{Instead of $\qr$ the nucleon momentum $p^2$ can be used as an alternative variable, but the final result for $C_{\vecq,\vecr}^{\rm corr}$ does not simplify for this choice.}
\begin{eqnarray}
C_{\vecq,\vecr}^{\rm corr} & = & \cc{0}+\cc{q^2}\qoe+\cc{r^2}\roe + \cc{qr}\qroe\nn\\
& & +\cc{q^2r^2}\qoe\roe+\cc{q^2qr} \qoe\qroe + \cc{r^2qr} \roe\qroe +\cc{(qr)^2} \qrqroe\nn\\
& & + \cc{q^4r^2} \qqoe\roe + \cc{q^2r^4} \qoe\rroe  + \cc{q^2r^2qr}\qoe\roe\qroe\nn\\
& & + \cc{q^2(qr)^2}\qoe\qrqroe +  \cc{r^2(qr)^2}\roe\qrqroe + \cc{(qr)^3}\qrqrqroe\,,
\end{eqnarray}
where the newly introduced coefficients read
\begin{eqnarray}
 \cc{0} &=& -\frac{\ga \left(3 \Epiq^2-2 \Epiq \Epir (\ga+1)+3
   \Epir^2\right)}{(\Epiq+\Epir)^2}\\
 \cc{q^2}&=&-\frac{2 \Epiq^2 \left(\Epiq^2 \ga^2-2 \Epiq \Epir
   \left(\ga^2+1\right)+\Epir^2 \left(2-3 \ga^2\right)\right)}{(\Epiq+\Epir)^4}\\
    \cc{r^2}&=&
   \frac{2 \Epir^2 \left(\Epiq^2 \left(3 \ga^2-2\right)+2 \Epiq \Epir
   \left(\ga^2+1\right)-\Epir^2 \ga^2\right)}{(\Epiq+\Epir)^4}\\
  \cc{qr} & =&\frac{(\Epiq^4+\Epir^4) \ga \left(\ga^2-\ga+1\right)+2 (\Epiq^3 \Epir+ \Epiq \Epir^3) \left(\ga^4-2 \ga^3+2 \ga^2-2
   \ga-2\right)}{(\Epiq+\Epir)^4}\nn\\
   & &+\frac{2\Epiq^2 \Epir^2 \left(2 \ga^4-5 \ga^3+5 \ga^2-5 \ga+4\right)}{(\Epiq+\Epir)^4}\\
\cc{q^2r^2} & = & \frac{ 3 (\Epiq^2+\Epir^2) (\ga-1) \ga^3+2 \Epiq \Epir
   (\ga^2-\ga-2)\ga^2}{(\Epiq+\Epir)^2}\\
   \cc{q^2qr}&=&\frac{\Epiq \ga^2 \left(\Epiq^3 \left(6 \ga^2+1\right)+\Epiq^2 \Epir \left(4
   \ga^2-13\right)+\Epiq \Epir^2 \left(3-2
   \ga^2\right)+\Epir^3\right)}{(\Epiq+\Epir)^4}\\
    \cc{r^2qr}&=&\frac{\Epir \ga^2 \left(\Epiq^3+\Epiq^2 \Epir \left(3-2 \ga^2\right)+\Epiq
   \Epir^2 \left(4 \ga^2-13\right)+\Epir^3 \left(6
   \ga^2+1\right)\right)}{(\Epiq+\Epir)^4}\\
   \cc{(qr)^2} & = & \frac{2 \ga^2 \left(\Epiq^4+\Epir^4-(\Epiq^3 \Epir + \Epiq \Epir^3)\left(2 \ga^2-2 \ga-3\right)-4
   \Epiq^2 \Epir^2 \left(\ga^2-\ga+3\right)\right)}{(\Epiq+\Epir)^4}\nn\\
   & &
 \end{eqnarray}
 \begin{eqnarray}  
 \cc{q^4r^2} & = & -\frac{\Epiq \ga^4 \left(3 \Epiq^3+3 \Epiq^2 \Epir+7 \Epiq \Epir^2+3
   \Epir^3\right)}{(\Epiq+\Epir)^4}\\
 \cc{q^2r^4}  & = & -\frac{\Epir \ga^4 \left(3 \Epiq^3+7 \Epiq^2 \Epir+3 \Epiq \Epir^2+3
   \Epir^3\right)}{(\Epiq+\Epir)^4}\\  
   \cc{q^2r^2qr} &=& -\frac{2 \ga^4 \left(3 \Epiq^4+7 \Epiq^3 \Epir+4 \Epiq^2 \Epir^2+7 \Epiq
   \Epir^3+3 \Epir^4\right)}{(\Epiq+\Epir)^4}\\
   \cc{q^2(qr)^2}&=& \frac{2 \Epiq^2 \Epir \ga^4 (3 \Epiq+\Epir)}{(\Epiq+\Epir)^4}\\
   \cc{r^2(qr)^2}&=&\frac{2 \Epiq \Epir^2 \ga^4 (\Epiq+3 \Epir)}{(\Epiq+\Epir)^4}\\
 \cc{(qr)^3} &=& \frac{4 \Epiq \Epir \ga^4 \left(\Epiq^2+4 \Epiq
   \Epir+\Epir^2\right)}{(\Epiq+\Epir)^4}
 \end{eqnarray}
The result simplifies significantly for some special cases. If both pions are at rest, only the $c_0$ term contributes to $C_{0,0}^{\rm corr} $ and we obtain the simple result given in \pref{resC00}. In case at least one pion is at rest only two terms (proportional to $c_0$ and either $c_{q^2}$ or $c_{r^2}$) provide a non-vanishing contribution.

\section{Degeneracies of the lowest 3-particle states}
\label{appB}

The sum in \pref{DefMeff} runs over the two pion momenta $\vec{q},\vec{r}$ that are allowed by the spatial boundary conditions. In case of periodic boundary conditions these are given in \pref{pimomenta} with integer-valued vectors $\vec{n}_q,\vec{n}_r$. To get a sum over a finite number of low-momentum pion states we impose an upper bound $\Lambda_{\rm max}$ on the absolute values of the pion momenta. Once the spatial extent $L$ is fixed this is done by imposing a bound $n_{\rm max}$ on the integers $n_q$ and $n_r$, defined in \pref{Defnq}. 

Symmetry under the $O_3$ group implies that the sum over the pion momenta $\vec{q},\vec{r}$ simplifies to a 3-fold sum over the discrete values for $|\vec{q}|,|\vec{r}|$ and the scalar product $\vec{q}\cdot\vec{r}$. Since the nucleon momentum $\vec{p}$ is fixed by momentum conservation, $\vec{p}=-\vec{q}-\vec{r}$, the scalar product can be replaced by $|\vec{p}|$. In that representation the momentum sum in \pref{DefMeff} is replaced by 
\begin{equation}
\sum_{\vec{q},\vec{r}}^{\Lambda_{\rm max}} = \sum_{n_{q},n_{r}}^{n_{\rm max}} \sum_{n_p} m(n_p)\,.
\end{equation}
Here $n_p$ runs over a finite number of allowed integers, determined by 
\begin{equation}\label{Defnp}
n_p=n_q+n_r+2\,\vec{n}_q\cdot\vec{n}_r.
\end{equation}
$m(n_p)$ gives the multiplicities of the particular three-particle state with given values $n_q,n_r$ and $n_p$, i.e.\ it counts the number of ways one can find three integer-valued vectors $\vec{n}_p,\vec{n}_q,\vec{n}_r$ with given values $n_p,n_q,n_r$ that add up to zero,  $\vec{n}_p+\vec{n}_q+\vec{n}_r=0$.

As long as $n_q,n_r$ are sufficiently small $n_p$ and $m(n_p)$  are straightforwardly computed by explicit calculations of \pref{Defnp}.
The results are summarized for $1\le n_q,n_r\le 8$ in tables \ref{tab:npmnp1} and \ref{tab:npmnp2} (note that no vector with $n_q,n_r=7$ exists).
In case either $n_q$ or $n_r$ equals zero one of the pions is at rest and the nucleon momentum is opposite to the non-zero pion momentum. The multiplicities for these special cases are the same as for the $N\pi$ contribution and are listed in Ref.\ \cite{Bar:2015zwa}, for example. The state with vanishing momenta for both pions is the state with all three particles at rest, which is non-degenerate.

\begin{table}[p]
\begin{center}
\begin{tabular}{c|c|rr}
$n_q$ & $n_r$ & $n_p$ & $m(n_p)$ \\ \hline
1 & 1 & 0 & 6\\
 & & 2 & 24\\
  & & 4 & 6\\ \cline{2-4}
  & 2 & 1 & 24 \\
  & & 3 & 24\\
  & & 5 & 24\\ \cline{2-4}
 & 3& 2 & 24\\
 & & 6 & 24\\ \cline{2-4}
 & 4 & 1 & 6\\
 & & 5 & 24\\
 & & 9 & 6\\ \cline{2-4}
 & 5 & 2 & 24\\
 & & 4&24\\
 & & 6&48\\
 & & 8&24\\
 & & 10& 24\\ \cline{2-4}
 & 6 & 3& 24\\
 & & 5&48\\
 & & 9 & 48\\
 & & 11 & 24\\ \cline{2-4}
 & 8 & 5& 24\\
 & & 9 & 24\\
 && 13 & 24\\\hline \hline
 \end{tabular}\hspace{1cm}
 \begin{tabular}{c|c|rr}
$n_q$ & $n_r$ & $n_p$ & $m(n_p)$ \\ \hline
2 & 2& 0&12\\
& & 2&48\\
& & 4 & 24\\
& & 6 & 48\\
& & 8 & 12\\\cline{2-4}
& 3 & 1 & 24\\
& & 5 & 48\\
& & 9 & 24\\ \cline{2-4}
& 4 & 2 & 24\\
& & 6 & 24\\
& & 10 & 24\\ \cline{2-4}
& 5 & 1 & 24\\
& & 3 & 48\\
& & 4 & 24\\
& & 5 & 48\\
& & 9 & 72\\
& & 11 & 48 \\
& & 13 & 24\\\cline{2-4}
& 6 &2&48\\
& & 4&24\\
& & 6 & 48\\
& & 8 & 48\\
& & 10 & 48\\
& & 12 & 24\\
& & 14 & 48\\\cline{2-4}
& 8 & 2 & 12\\
& & 6 & 48\\
& & 10 & 24\\
& & 14 & 48\\
& & 18 & 12\\
\hline\hline
\end{tabular}\hspace{1cm}
\begin{tabular}{c|c|rr}
$n_q$ & $n_r$ & $n_p$ & $m(n_p)$ \\ \hline
3&3&0&8\\
& & 4 & 24\\
& & 8&24\\
& &12& 8\\ \cline{2-4}
& 4&3&24\\
& & 11&24\\ \cline{2-4}
& 5 & 2&48\\
& & 6&48\\
& & 10&48\\
& & 14&48\\ \cline{2-4}
& 6 & 1 & 24\\
& & 5 & 48 \\
& & 9 & 48\\
& & 13 & 48\\
& & 17 & 24\\\cline{2-4}
& 8 & 7 & 24\\
& & 11& 48\\
& & 15 & 24\\
\hline\hline 
4 & 4 & 0 & 6\\
& & 8 & 24\\
& & 16 & 6\\ \cline{2-4}
& 5 & 1 & 24\\
&  & 5 & 24\\
& & 9 & 48\\
& & 13 & 24\\
& & 17 &24\\ \cline{2-4}
& 6 & 2 & 24 \\
& & 6 & 48\\
& & 14&48\\
& & 18 & 24\\\cline{2-4}
& 8 & 8 & 24\\
& & 12 & 24\\
& & 16 & 24\\
\hline\hline
\end{tabular}
\end{center}
\caption{Possible values for the nucleon momentum integer $n_p$ and the multiplicities $m(n_p)$ as a function of $n_q,n_r$.}
\label{tab:npmnp1}
\end{table}%

\begin{table}[p]
\begin{center}
\begin{tabular}{c|c|rr}
$n_q$ & $n_r$ & $n_p$ & $m(n_p)$ \\ \hline
5 & 5 & 0 & 24\\
&& 2& 72\\
&& 4&24\\
&& 6 & 96\\
&& 8 & 48\\
&&10 & 48\\
&& 12 &48\\
&& 14 &96\\
&& 16 &24\\
&& 18 & 72\\
&&20 & 24\\\cline{2-4}
& 6 & 1 & 48\\
& & 3 & 48\\
& & 5 & 96\\
& & 9 & 48\\
& & 11 & 96\\
& & 13 & 48\\
& & 17 & 96\\
& & 19 & 48\\
& & 21 & 48\\\cline{2-4}
& 8 & 1 & 24\\
& & 5 & 48\\
& & 9 & 24\\
& & 11&48\\
& & 15 & 48 \\
& & 17 & 24 \\
& &21 & 48\\
& & 25 & 24\\
\hline\hline
\end{tabular}\hspace{1cm}
\begin{tabular}{c|c|rr}
$n_q$ & $n_r$ & $n_p$ & $m(n_p)$ \\ \hline
6 & 6 & 0 & 24\\
& & 2 & 48\\
& & 4 & 48\\
& & 6 & 48\\
& & 8 & 24\\
& & 10 & 96\\
& & 14 & 96\\
& & 16 & 24\\
& & 18 & 48\\
& & 20 & 48\\
& &22 & 48 \\
& &24 & 24\\\cline{2-4}
& 8 & 2 & 48 \\
& & 6 & 24 \\
& & 10 & 48 \\
& & 14 & 48 \\
& & 18 & 48 \\
& & 22 & 24\\
& & 26 & 48\\
 \hline\hline 
 8 & 8 & 0 & 12\\
& & 8 & 48\\
& & 16 & 24\\
& & 24 & 48\\
& & 32 & 12\\ \hline\hline
\end{tabular}
\end{center}
\caption{Possible values for the nucleon momentum integer $n_p$ and the multiplicities $m(n_p)$ as a function of $n_q,n_r$ (cont.)}
\label{tab:npmnp2}
\end{table}%

\end{appendix}

\end{document}